\documentstyle[twoside,fleqn,espcrc2]{article}

\newcommand{\AmS}{{\protect\the\textfont2
  A\kern-.167em\lower.5ex\hbox{M}\kern-.125emS}}

\title{Anisotropic Improved Gauge Actions:\\
 -- Perturbative and Numerical Studies --}
        
\author{S. Sakai,\address{Faculty of Education, Yamagata University\\},
     A. Nakamura,\address{Research Center for Information Science 
       and Education, Hiroshima University\\}, and
T.Saito,\address{Department of Physics, Hiroshima University}
       }

\begin{document}

\begin{abstract}
The $\Lambda$ parameter on the anisotropic lattice, the spatial and 
temperature coupling constant $g_{\sigma}$, $g_{\tau}$ and their 
derivative with respaect to the the anisotropy parameter $\xi$ are
studied perturbatively for the class of improved actions, which
cover tree level Symanzik's, Iwasaki's and QCDTARO's improved actions.
The $\eta(=g_{\tau}/g_{\sigma}) $ becomes less than $1$ 
for Iwasaki's and QCDTARO's action,
which is confirmed nonperturbatively by numerical simulations.
Derivatives of the coupling constants with respect to the anisotropy parameter
, $\partial g_{\tau}/\partial \xi$ and 
$\partial g_{\sigma}/\partial \xi$, change sign for those 
improved actions.
\end{abstract}

\maketitle

\section{INTRODUCTION}
The basic parameters of anisotropic lattices
with improved action are important information
for numerical simulations.
In this work, we mainly focus on perturbative
calculations of these parameters, 
but some of them are confirmed by 
numerical simulations.\\ 
\indent
The improved actions we study in this work consist of terms
$$ S_{\mu \nu} = \alpha_{0} \cdot L(1\times 1)_{\mu \nu}
 + \alpha_{1} \cdot L(1\times2)_{\mu \nu}$$
where $L(1\times 1)$ and $L(1\times 2)$ represent plaquette and 6-link
rectangular loops respectively, and $\alpha_{0}$ and $\alpha_{1}$
satisfies $\alpha_{0}+8 \cdot \alpha_{1}=1$.
The improved actions cover 
i) tree level Symanzik's action($\alpha_{1}=-1/12$)\cite{Symanzik},
ii) Iwasaki's action($\alpha_{1}=-0.331$)\cite{Iwasaki}, and
iii) QCDTARO's action($\alpha_{1}=-1.409$)\cite{Taro} etc.\\
\indent
For these classes of improved actions the anisotropic lattice is formulated
in the same way as in the case of standard plaquette action\cite{Karsch}.
We introduce the coupling constant and lattice spacing 
in space direction $g_{\sigma}$, $a_{\sigma}$ and those in
temperature direction $g_{\tau}$, $a_{\tau}$.
With these parameters the action on the anisotropic lattice is
written as,
$$ S = \beta_{\sigma} \cdot S_{ij} + \beta_{\tau} \cdot S_{4i}$$
where
$\beta_{\sigma}=g_{\sigma}^{-2} \xi_{R}^{-1}$,
$\beta_{\tau}=g_{\tau}^{-2} \xi_{R}$ and 
$\xi_{R}=a_{\sigma}/a_{\tau}$.\\  
\indent
In this study,  
$\Lambda$ parameter and the ratio
$\eta = g_{\tau}/g_{\sigma}$ are calculated and the derivatives
of the coupling constants with respect to the anisotropy parameter, 
$\partial g_{\sigma}^{-2}/\partial \xi_{R}$,
$\partial g_{\tau}^{-2}/\partial \xi_{R}$
will be discussed.

\section{PERTURBATIVE CALCULATIONS}

In weak coupling expansions,
the relation between the couping constants on the anisotropic
lattice and those on the isotropic lattice are obtained
as follows\cite{Karsch}.
We calculate the coefficients of $F_{ij}^{n}F_{ij}^{n}$ and 
$F_{i4}^{n}F_{i4}^{n}$ in the action in one loop order.
\begin{eqnarray*} \lefteqn{S_{Eff}(\xi_{R})}\\
 &=& \frac{1}{4}(g_{\sigma}^{-2} -C_{\sigma}(\xi_{R})) \cdot
F_{ij}^{n}F_{ij}^{n} \cdot a_{\sigma}^{3}a_{\tau}\\
 &+& \frac{1}{4}(g_{\tau}^{-2} -C_{\tau}(\xi_{R}))
\cdot F_{4i}^{n}F_{4i}^{n} \cdot a_{\sigma}^{3}a_{\tau}\\
\end{eqnarray*} 
The same calculation is done on the isotropic lattice. 
We require that the effective action is independent of
the method of regularization. Then the relations are derived.
$$g_{\sigma}^{-2}(\xi_{R})=g(1)^{-2}+(C_{\sigma}(\xi_{R})-C_{\sigma}(1))
+O(g(1)^{2}) $$
$$g_{\tau}^{-2}(\xi_{R})=g(1)^{-2}+(C_{\tau}(\xi_{R})-C_{\tau}(1))
+O(g(1)^{2})$$ 
On an isotropic lattice($\xi_{R}=1$), gauge 
invariance means $C_{\sigma}(1)=C_{\tau}(1)$.\\
\indent
For the calculation of $C_{\sigma}$ and $C_{\tau}$, we employ
the background field method\cite{Dashen}
used by Iwasaki and Sakai\cite{Sakai}
in the calculation of the $\Lambda$ parameters of improved actions
on the isotropic lattice.
The details of the method will be reported in the forthcoming 
publications.\\
\indent
The $C_{\tau}(\xi_{R})$ and 
$C_{\sigma}(\xi_{R})$ have infrared divergences, but in the difference
they are cancelled. The cancellation of the divergence is a delicate
problem. The following method is used.  
\begin{eqnarray*} 
\lefteqn{\Delta C_{\sigma}(\xi_{R})=
C_{\sigma}^{Imp}(\xi_{R})-C_{\sigma}^{Imp}(1)} \\
 & =&  (C_{\sigma}^{Imp}(\xi_{R})-C_{\sigma}^{Stand}(\xi_{R}))\\
 &+& (C_{\sigma}^{Stand}(\xi_{R})-C_{\sigma}^{Stand}(1))\\
 &+& (C_{\sigma}^{Stand}(1)-C_{\sigma}^{Imp}(1))\\
\end{eqnarray*}
In the first and third terms, the infrared divergence is canceled
exactly for improved and standard actions.
The second term is already calculated by 
Karsch\cite{Karsch} where the regularization of the infrared divergence
could be done by an analytic integration of the fourth component of
the loop momentum.\\
\indent
With these $C_{\sigma}(\xi_{R})$,$C_{\tau}(\xi_{R})$ functions, 
$\Lambda$ and $\eta$ are expressed as follows.
$$\Lambda(\xi_{R})/\Lambda(1)=
exp(-\frac{\Delta C_{\sigma}(\xi_{R})+\Delta C_{\tau}(\tau)}{4b_{0}})$$
$$\eta = \frac{g_{\tau}}{g_{\sigma}}=1+N \cdot (C_{\sigma}-C_{\tau})/\beta
+O(g^{4})$$  
The ratios $\Lambda(\xi_{R})/\Lambda$ are shown in Fig.1. 
\begin{figure}[htb]
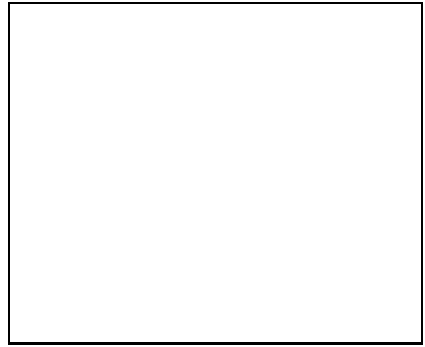

\vspace{9pt} 
\framebox[55mm]{\rule[-21mm]{0mm}{43mm}}
\caption{The ratio $\Lambda(\xi_{R})/\Lambda(1)$ for 
improved and standard actions}
\label{fig:largenenough}
\end{figure}
Except for the QCDTARO's action,
they have similar $\xi_{R}$ dependences.
The more interesting differences are seen in the ratio $\eta$. 
In Fig.2, we show $C_{\sigma}(\xi_{R})-C_{\tau}(\xi_{R})$,
which determines $\eta$.
\begin{figure}[htb]
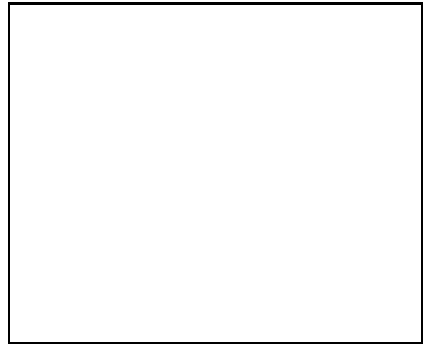

\vspace{9pt}
\framebox[55mm]{\rule[-21mm]{0mm}{43mm}}
\caption{$C_{\sigma}(\xi_{R})-C_{\tau}(\xi_{R})$ as a 
function of $\alpha_{1}$ }
\label{fig:largenenough}
\end{figure}
It is found that as $\alpha_{1}$ decreases
$C_{\sigma}(\xi_{R})-C_{\tau}(\xi_{R})$ changes sign. 
Namely for Iwasaki's and QCDTARO's 
improved actions, $\eta < 1$,
and at $\alpha_{1} \sim -0.160$, 
$C_{\sigma}(\xi_{R})-C_{\tau}(\xi_{R}) \sim 0$ for all $\xi_{R}$. 
\indent
This result is very interesting and 
is not what we have expected.
Then in the next section, we calculate the parameter $\eta$ 
nonperturbatively and confirm the perturbative results.

\section{NONPERTURBATIVE STUDY OF $\eta$}
Nonperturbatively, the parameter $\eta$ is calculated by the
relation $\eta = \xi_{R}/\xi_{B}$\cite{Klassen}\cite{Nakamura}\cite{Tim}
, where $\xi_{B}$ is a bare
the anisotropy parameter which appears in the action,
$$ S = \frac{2N}{g^{2}}(\frac{1}{\xi_{B}}S_{ij}+\xi_{B}S_{i4})$$
while the renormalized the anisotropy parameter is defined by 
$\xi_{R}= a_{\sigma}/a_{\tau}$.  For the probe of the scale in the
space and temperature direction, we use the lattice potential 
in these directions
which is defined by,
$$V_{st}(\xi_{B},l,t)=ln(\frac{W_{st}(l,t)}{W_{st}(l+1,t)})$$
similarly for the potential in the space direction.
Here we fix $\xi_{R}=2$, and calculate the rate for a few $\xi_{B}$
$$R(\xi_{B},l,r)=\frac{V_{ss}(\xi_{B},l,r)}{V_{st}(\xi_{B},l,\xi_{R}t)}$$
Then we search for the point $R=1$ by interpolating 
with respect to
$\xi_{B}$\cite{Klassen}\cite{Nakamura}.\\
\indent
The simulations are done 
on the $12^{3} \times 24 $ lattice, 
in the large $\beta$ region.
The results are shown in Fig.3, together with those of 
perturbative calculations.\\ 
\begin{figure}[htb]
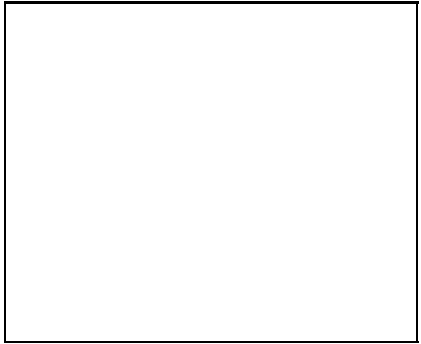

\vspace{9pt} 
\framebox[55mm]{\rule[-21mm]{0mm}{43mm}}
\caption{Perturbative and Nonperturbative results for $\eta(\xi_{R}=2,\beta)$}
\label{fig:largenenough}
\end{figure}
\indent
It is found that the agreement of both results is good.
So the $\alpha_{1}$ dependences of the $\eta$ ratio is also confirmed
by the numerical simulation, a very
exciting result for improved actions.
\section{ESTIMATION OF 
$\frac{\partial g_{\sigma}^{-2}}{\partial \xi_{R}}$
AND $\frac{\partial g_{\tau}^{-2}}{\partial \xi_{R}}$ }

The exact calculation of these derivatives are not carried out yet.
However we could estimate them from the slope of the 
$C_{\sigma}(\xi_{R})$ and $C_{\tau}(\xi_{R})$ at $\xi_{R}=1$ 
and $\xi_{R}=1.01$.
The results are shown in Fig.4. 
\begin{figure}[htb]
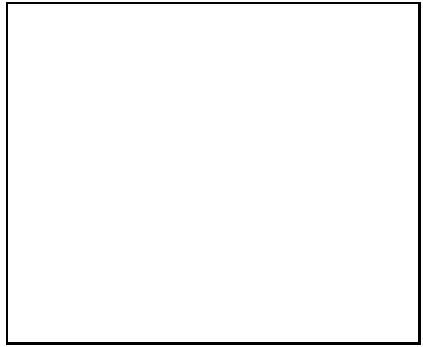

\vspace{9pt} 
\framebox[55mm]{\rule[-21mm]{0mm}{43mm}}
\caption{$\partial g_{\sigma}^{-2}/\partial \xi_{R}$,
and $\partial g_{\tau}^{-2}/\partial \xi_{R}$ from numerical
estimation}
\label{fig:largenenough}
\end{figure}
They should satisfy following sum rule\cite{Karsch},
$$ \frac{\partial C_{\sigma}}{\partial \xi_{R}} + 
\frac{\partial C_{\tau}}{\partial \xi_{R}}|_{\xi_{R}=1}= 
\frac{11 \times N}{48 \pi^{2}}$$
The saturation of the sum rule is good.\\
\indent
Notice that the derivatives change sign as $\alpha_{1}$ decreases.
These coefficients play important role in the calculation of 
internal energy and pressure.
As the sign of these coefficients changes for the
improved actions, it may be interesting to calculate
thermodynamic quantities with these improved actions.\\
\indent
In conclusion, the study of anisotropic lattice with improved action 
shows  
very interesting properties as shown in Figs.1,2,3,4. The
study of finite temperature physics with these improved
action seems to be very exciting. 


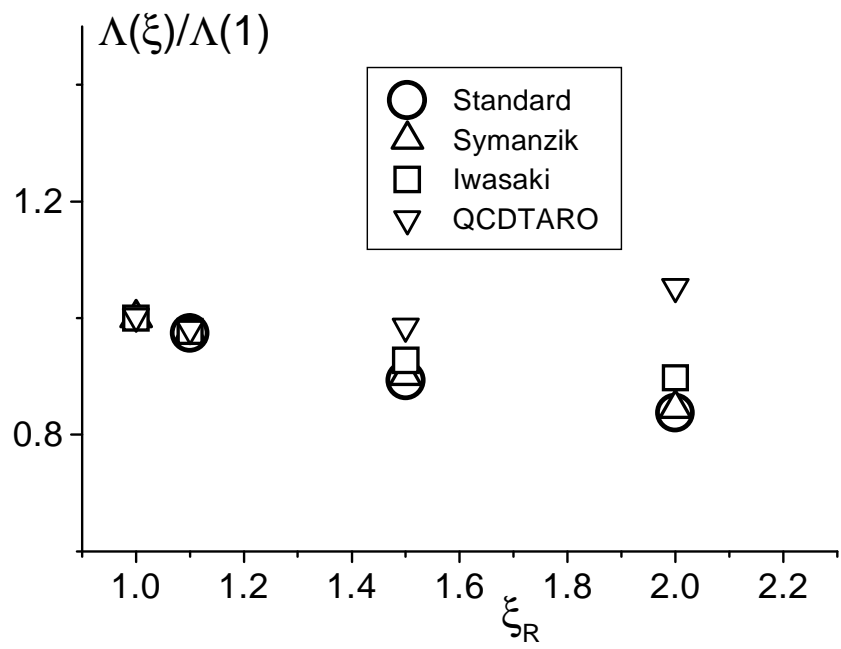
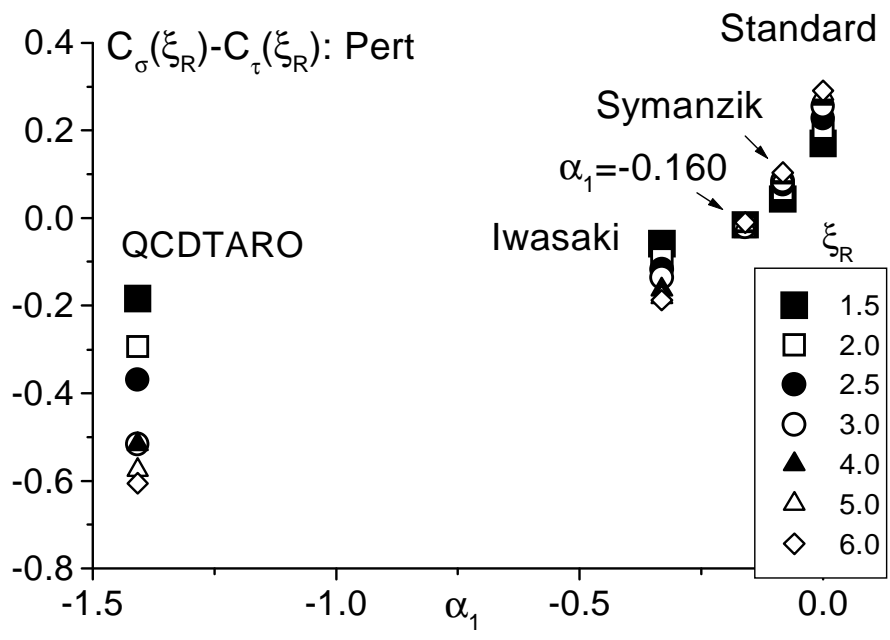
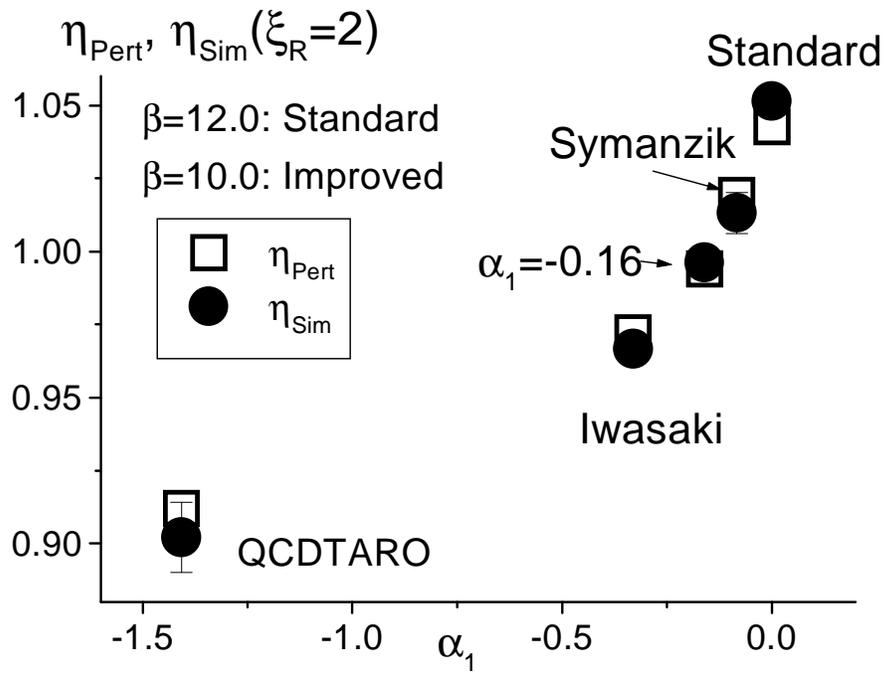
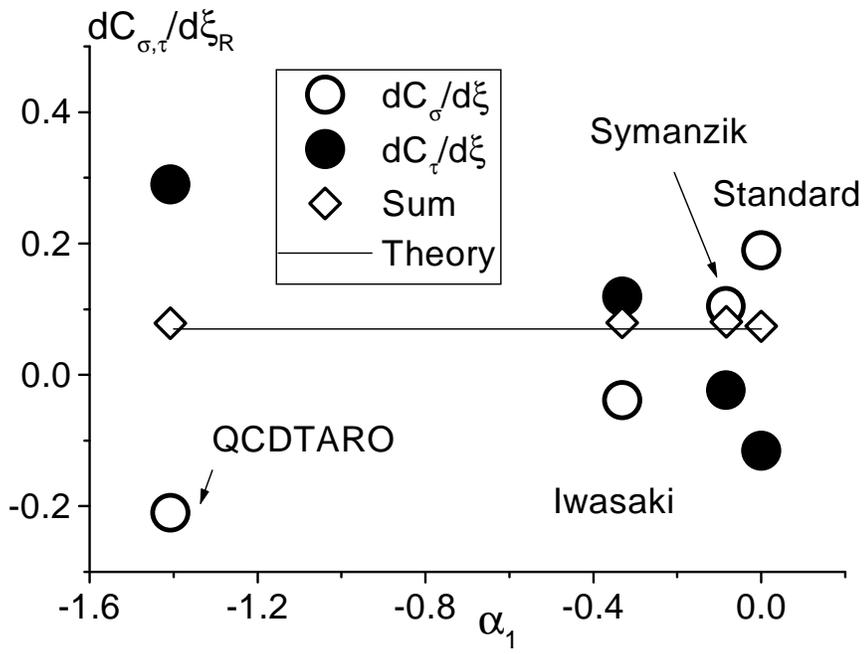
\end{document}